\documentclass{aastex}          
\usepackage{graphicx} 
\usepackage{spr-astr-addons}    
\usepackage{url}

\begin{document}
\title{The intermediate age population of the Galactic halo}
\author{Jeremy Mould$^{1}$}
\altaffiltext{1}{ Centre for Astrophysics and Supercomputing, Swinburne University of Technology, Melbourne, Victoria 3122, Australia}

\footnote{jmould@swin.edu.au}




\begin{abstract}
We have learned recently that the inner halo of the Milky Way contains a kinematically coherent
component (Gaia-Enceladus) from a significant merger 10 Gyrs ago. By contrast the inner\footnote{defined to exclude the Magellanic Stream} halo
contains no similar intruder stellar population of billion year age. The tracer we use to set the
corresponding upper limit is Gaia asymptotic giant branch stars, rather than Gaia kinematics. 
The primary sample is drawn from Gaia DR2 with SkyMapper photometry.
This is supplemented with PanSTARRS and 2MASS photometry.
As the Gaia mission proceeds, a star formation history in the galactic halo should
emerge.

\end{abstract}
\begin{keywords}
 .Parallax, Stars -- red giants, galaxies: halos 
\end{keywords}

\maketitle
\section{ Introduction}
Research into the halo of our Galaxy is currently extraordinarily fruitful. The ESA $Gaia$ satellite promises distances and kinematics of most of the stars in the halo. Halo streams, of which the first was the Sgr dwarf (Majewski et al 1999), are being mapped by deep wide field surveys (Shipp et al 2018, Helmi et al 2018). Spectroscopy of Southern streams includes the populous main sequence (Li et al 2018). Distant and diffuse components are being detected with the aid of kinematic filtering (Torrealba et al 2018). Antlia 2 is just one of
many at large distances (see Appendix), and much more is to come (Malhan
et al 2019).
Inhomogeneities in the dark matter halo are calculable, vital for our understanding of dark matter and targets for direct detection experiments (Lawrence 2018). Deep imaging of external galaxies are yielding rich returns (Abraham et al 2017).

The goal, as it has been for 50 years, is to understand the formation of the halo. Accretion of dwarf galaxies is the paradigm for the outer halo, and the inner bulge is evidently very old (Carollo et al 2018) and may originate differently. The star formation history of the halo is part of this story and it is what we focus on here. Intermediate age stars are seen in the Magellanic Clouds, but not in the halo of the Milky Way. We endeavour to quantify that contrast here using parallax data from Data Release 2 of $Gaia$.

Long period variable stars have been used to sketch the star formation history of Local Group dwarf galaxies in a similar way (Saremi et al 2018).
\section{Asymptotic Giant Branch Sample}
Our sample for this purpose is a stellar population drawn from the thick disk \& halo of the Milky Way.  
\subsection{Database Query}
The input catalogs used here are Data Release DR1.1 of the SkyMapper survey of the southern sky (Wolf et al. 2018), which incorporates a crossmatch to Gaia DR2 catalog, and the PanSTARRS Gaia catalog in the ESA $Gaia$ archive.

\vfill\break
In the former case this led to the following database query:

{\it \noindent select top 250000 raj2000, dej2000,}\\
{\it i$\_$psf, e$\_$i$\_$psf, i$\_$nimaflags, bp$\_$rp,ebmv$\_$sfd,\\
parallax, parallax$\_$error, z$\_$psf, astrometric$\_$excess$\_$noise\\}
{\it FROM DR1.master m join ext.gaia$\_$dr2 on \\
(source$\_$id=m.gaia$\_$dr2$\_$id1 and }\\
{\it m.gaia$\_$dr2$\_$dist1$<$1) WHERE (glat$<$=-10 OR glat$>$10)\\
 AND (phot$\_$r$\_$mean$\_$mag$<$=12 AND bp$\_$rp$>$=1}\\
{\it   AND parallax $>$= .1 AND i$\_$psf$>$=7)}
\\

The southern colour magnitude diagram (CMD) (Figure 1) is corrected for reddening following Schneider et al (1983) and Schlegel et al (1998),
and a bolometric correction (BC) is applied to the average of the $i$ and $z$ magnitudes, based on tables\footnote{http://pleiadi.pd.astro.it} calculated by 
Girardi et al. (2004). This BC varies slowly from --0.05 to 0.30 mag,
as $i-z$ changes from 0.50 to 0.23 mag.
Modifications to the query, changing the magnitude limit from 12 to 11.5, make
no significant difference.
The parallax distribution for stars with M$_{bol}~<$ --2.5 mag is shown in Figure 2.
Figure 3 shows the distribution of our sample in celestial coordinates, centred on the South Galactic Pole (SGP). Gaia DR2 mean parallax offsets are discussed by Arenou et al (2018) and Chen et al (2018). 

\section{Analysis}
\subsection{The brightest stars}

Adopting a single mean parallax offset for the entire sample, we collected
stars with M$_{bol}~<$ --4.5 mag in Table 1. It is noteworthy that the average height above/below the plane of these stars is 1.5 kpc, suggesting a large thick disk contribution, a subject we return to in $\S4$.
This can be studied using Gaia kinematics, exemplified by Antlia 2 (see Appendix).

\begin{table*}[h]
\centering
\label{my-label}
\caption{Brightest halo AGB stars}
\begin{tabular}{rrrrccccrrcccc}
\hline
\textbf{\#} &n&RA&Dec$^\dagger$&Bp-Rp&$\pi$
&$\sigma$&M$_{bol}$&$i_{PSF}$&$z_{PSF}$&axn&J&H&K \\ 
&&(deg)&(deg)&(mag)&(mas)&(mas)&(mag)&(mag)&(mag)&&(mag)&(mag)&(mag)\\
\hline
 85605&     24& 124.8154& -60.1668&     1.15&     0.13&     0.03&    -5.02&     9.17&     9.09&     0.00& 8.81&7.72&6.50\\
 99682&     12& 104.3517& -27.8023&     2.76&     0.15&     0.03&    -4.66&     9.87&     9.22&     0.13& 7.38&6.32&5.74\\
108294&      6&  93.5231& -18.2217&     2.23&     0.20&     0.04&    -4.58&     9.22&     8.76&     0.00& 7.30&6.43&6.15\\
446680&     64& 103.9980& -51.2501&     1.98&     0.13&     0.02&    -5.05&     9.14&     8.89&     0.00& 7.40&6.63&6.33\\
458497&     19&  99.1644& -16.9928&     2.58&     0.19&     0.05&    -4.79&     8.85*&     8.47&     0.00& 6.21&5.20&4.64\\
467831&    112& 118.9336&  -3.1982&     2.57&     0.19&     0.04&    -4.62&     8.80*&     8.55&     0.00& 6.13&5.13&4.57\\
127532&      0&  26.1379&  10.9085&     1.97&     0.28&     0.06&    -5.28&     -&     -&     0.00&     5.53&     4.86&     4.52\\
179695&      0& 137.9400&  48.4241&     2.78&     0.26&     0.06&    -4.85&     -&     -&     0.07&     6.19&     5.26&     4.98\\
180822&      0& 229.6941&  30.8298&     2.93&     0.20&     0.05&    -4.64&     -&     -&     0.15&     6.93&     6.06&     5.79\\
182811&      0& 196.2554&  45.5090&     2.26&     0.16&     0.03&    -4.57&     -&     -&     0.00&     7.51&     6.71&     6.47\\
183000&      0& 267.6022&  23.1999&     2.72&     0.14&     0.04&    -4.63&     -&     -&     0.13&     7.68&     6.76&     6.52\\
219513&      0& 189.2233&  57.9688&     2.34&     0.20&     0.03&    -4.52&     -&     -&     0.00&     7.04&     6.21&     5.91\\
220292&      0& 325.7632&  82.9701&     2.43&     0.14&     0.03&    -4.76&     -&     -&     0.00&     7.96&     6.90&     6.39\\ 
\hline

\multicolumn{11}{l}{$Notes$: n is the \# of SkyMapper $i$ flagged pixels, i$\_$nimaflags, in the query}\\
\multicolumn{11}{l}{axn is the astrometric excess noise in the GAIA DR2 database}\\
\multicolumn{11}{l}{$\pi$ is the offset corrected parallax and $\sigma$ is parallax error}\\
\multicolumn{11}{l}{M$_{bol}$ is bolometric magnitude; Bp-Rp is the Gaia colour.}\\
\multicolumn{11}{l}{*Possibly slightly reddened in $i-z$ by saturation in $i$ band.}\\
\multicolumn{11}{l}{$^\dagger$Epoch 2000}
\end{tabular}
\end{table*}

\begin{figure}
\begin{center}
\includegraphics[width=1.6\columnwidth,angle=-90]{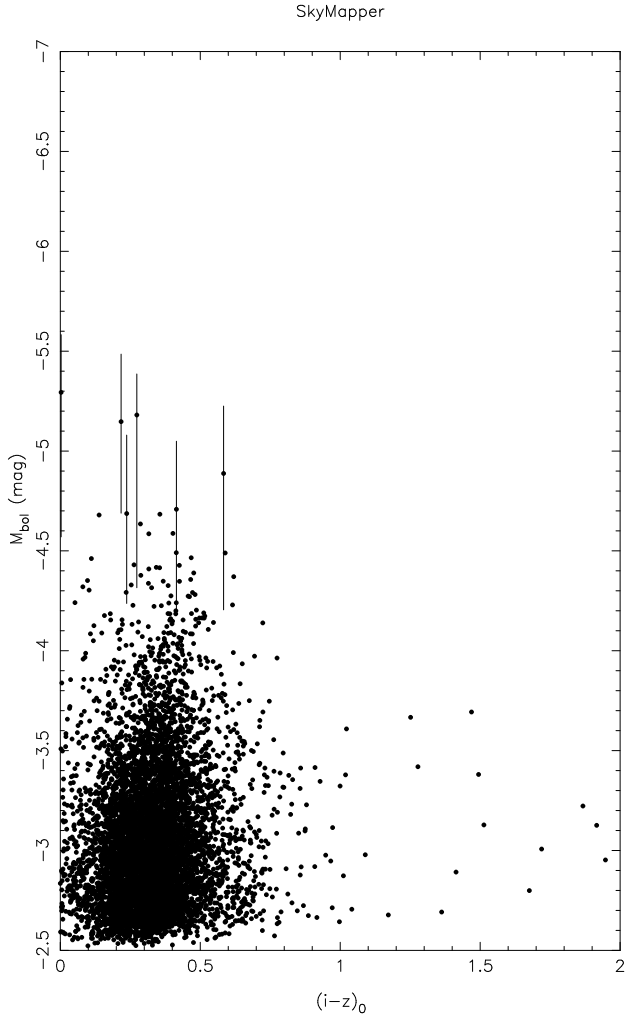}
\caption{SkyMapper Gaia DR2 red giant branch. Parallax errors are shown for the brightest stars. The colour is reddening corrected following Schneider, Gunn \& Hoessel (1983).
A Gaia DR2 mean parallax offset of 0.028 $\pm$ 0.006 mas was added to the data from Table 1 of Arenou et al (2018)
, appropriate to the Sculptor galaxy, which is at the South Galactic Pole.}
\end{center}
\end{figure} 

\begin{figure*}
\begin{center}
\includegraphics[width=\columnwidth,angle=-90]{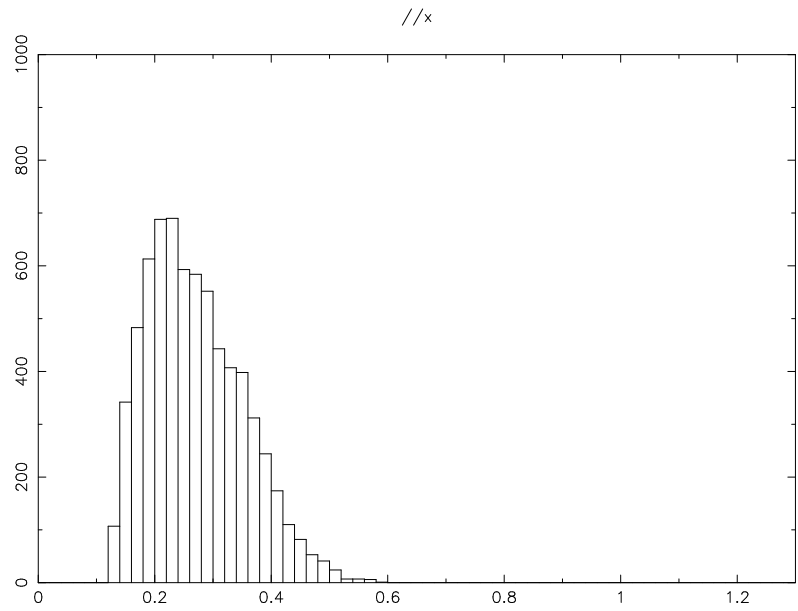}
\caption{The parallax distribution in mas for stars with M$_{bol}~<$ --2.5 mag in the SkyMapper-Gaia sample.}
\end{center}
\end{figure*} 
\begin{figure*}
\begin{center}
\includegraphics[width=1.05\columnwidth,angle=-90]{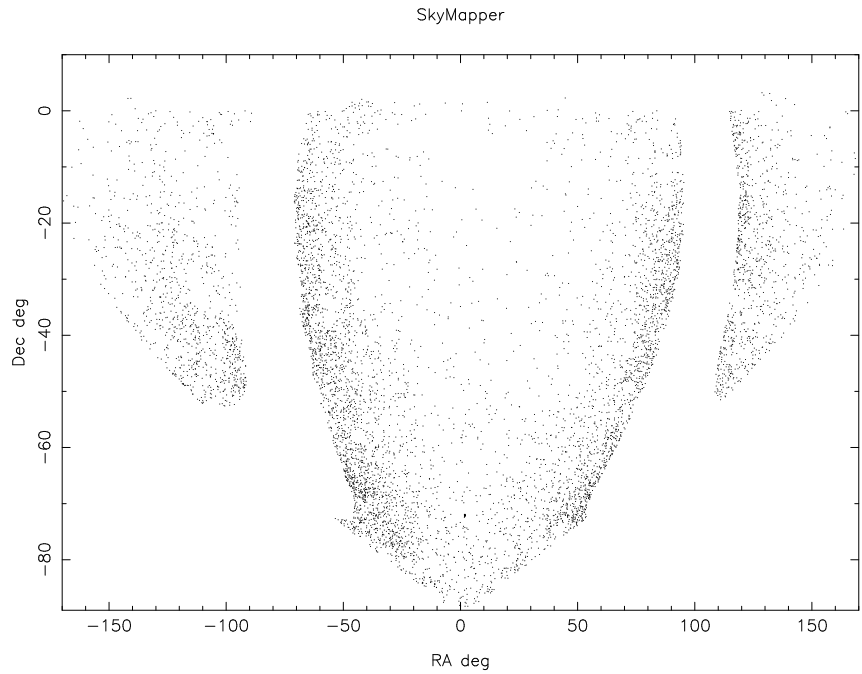}
\caption{Sky distribution of stars in Figure 1. }
\end{center}
\end{figure*} 

The corresponding CMD for a similar PanSTARRS database query on the ESA Gaia archive is shown in Figure 4. There is considerably
more scatter in $i-z$. The ESA Gaia archive was therefore used to query Gaia DR2 matched to 2MASS, obtaining Figures 5, 6 \& 7.
The separation between halo and thick disk was made at a distance from the plane of 2.2 kpc.
The separation between thin and thick disk was made 600 pc from the plane.
Bolometric corrections for 2MASS were taken from Frogel, Persson \& Cohen (1980). No reddening corrections were made.

As a control the DR2 query was modified to find AGB stars in the Large Magellanic Cloud, where the
intermediate age population is large.
Stars counted in these queries are summarized in Table 2.
\vspace*{0.5 truecm}
\begin{tabbing}
sssssSamplesssssssssss\=sssssssAGBsss\=RGB\kill
\small{\bf Table 2}\\
Gaia DR2\>AGB\>RGB\\
\>M$_{bol}<$--4.5\>--4$<$M$_{bol}<$--3.5\\
SkyMapper\>6*\>354\\
Halo 2MASS\>10*\>187\\
Thick disk 2MASS\>181*\>1250\\
LMC SkyMapper\>730\>2572\\
*\small{ Parallax uncertainty bias is likely to make this} \\
\small{ an upper limit. (See Mould et al 2018).}\\
\end{tabbing}

\subsection{Star Formation History}
Supposing the star formation history is composed of $j$ epochs, each of duration a Gyr, M$_j$. These simple stellar population masses, M$_j$, evolve to luminosities L$_j$.
According to Renzini \& Buzzoni (1986) 
the number of stars in post main sequence phase j is n$_j$ = B(t) L$_T$ $\Delta$t$_j$, 
where B(t) is the specific evolutionary flux (their figure 9), L$_T$ is the total luminosity of the simple stellar population (e.g. a star cluster) and $\Delta$t$_j$ is the duration of that phase. 
For the AGB population in the interval (--4.5,--5.5) mag in Figure 5 and Table 2 we have 10~=~B(t$_1$)L$_1$~1.3 where times are in Myrs and where $\Delta$t$_1$ is 1.3 Myr per mag according to the Paczynski core mass luminosity relation (Reid et al 1990, Renzini 1977). For the half mag below the RGB tip we have 187~=~B(t$_{13}$)L$_{13}$~1.7, using the pre-He flash red giant lifetimes of Rood (1972). 
If these are the only populations in this simplified model, 
the ratio L$_1$/L$_{13}$ = 10/1.3~1.7/187~B(t$_{13}$)/B(t$_1$) = 0.1 
From M/L (figure 4 of the same reference) we obtain M$_1$/M$_{13}~ \approx$ 1 $\times$ 10$^{-2}$ for t$_1~\approx$ 1 Gyr. The corresponding calculation for the SkyMapper sample in Figure 1 is M$_1$/M$_{13}~ \approx$ 3 $\times$ 10$^{-3}$. 
The latter sample is a hemisphere, and so Poisson statistics apply to the AGB counts in Table 2, rendering the
two calculations consistent.
It therefore seems best to express the result as a range from 0.3 -- 1\%. We neglect the systematic uncertainty 
of an assumed Salpeter initial mass function (IMF) on the grounds that, whatever may be the case for metal poor
halo populations,  there is no evidence that intermediate age populations have other than Salpeter IMFs on average. 

\begin{figure}
\begin{center}
\includegraphics[width=1.25\columnwidth,angle=-90]{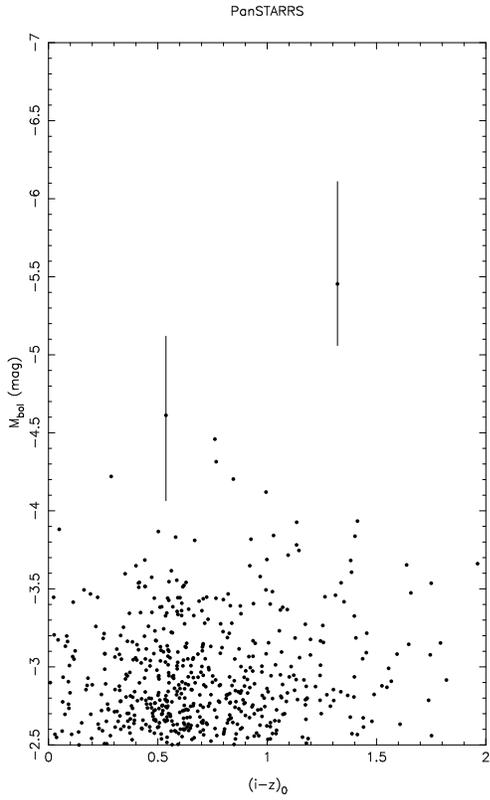}
\caption{The CMD from Gaia DR2 with PanSTARRS photometry.}
\end{center}
\end{figure} 
\begin{figure}
\begin{center}
\includegraphics[width=1.25\columnwidth,angle=-90]{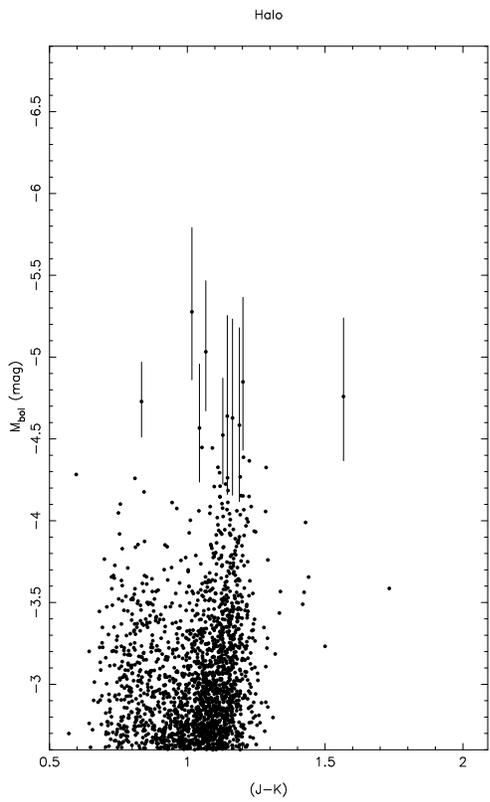}
\caption{The halo CMD from Gaia DR2 with 2MASS photometry.}
\end{center}
\end{figure} 
\begin{figure}
\begin{center}
\includegraphics[width=1.6\columnwidth,angle=-90]{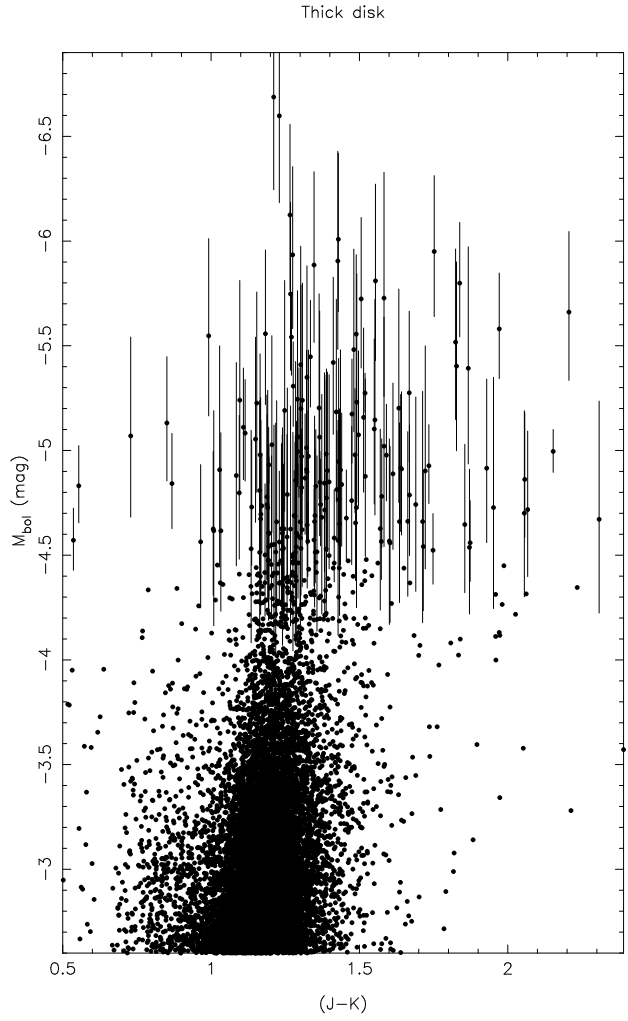}
\caption{The thick disk CMD}
\end{center}
\end{figure} 

\vfill\break
\begin{figure}
\begin{center}
\includegraphics[width=.75\columnwidth,angle=-90]{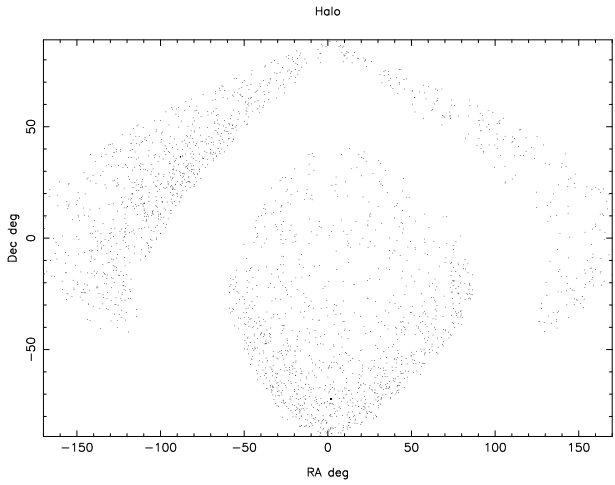}
\caption{Sky distribution for stars in Figure 5. Zero degrees RA is 12 hours.}
\end{center}
\end{figure} 

\section{Summary and future work}
The halo of the Milky Way within 6 kpc of the Sun is almost devoid of an intermediate age 
population. The upper limit on the mass ratio of the billion year old population to the older-than-10-Gyr
population is 1\%. Simple calculations based on Figure 6 and Table 2 suggest that this upper limit would be 
consistent with zero, were a correction for thick disk contamination to be applied.

This search can be improved as follows
\begin{itemize}
\item simulations of parallax uncertainty luminosity bias using a Galactic structure model
\item extending the volume surveyed to $\sim$30 kpc in future Gaia data releases
\item reducing the uncertainty in parallaxes and the parallax offset in these data releases
\item applying a Galactic structure model such as $Trilegal$ (Girardi 2016) would remove the tail of thick disk stars
\end{itemize}

It is interesting to compare this property of the Milky Way with the star formation history of M31.
According to Ferguson (2018) and Brown et al (2008) the accreted streams in M31's halo are predominantly old
(but surprisingly metal rich), like those of our Galaxy.
This contrasts with the disk, whose last major star forming event was $\sim$4 Gyr ago (Williams et al 2017)
and with the bulge, whose properties are summarized by Mould (2014). A study of M31's disk AGB stars
has been carried out by Chen et al (2018) and it would be interesting to extend this to the halo.

\vspace*{.5 truecm}
The national facility capability for SkyMapper has been funded through ARC LIEF grant LE130100104 from the Australian Research Council, awarded to the University of Sydney, the Australian National University, Swinburne University of Technology, the University of Queensland, the University of Western Australia, the University of Melbourne, Curtin University of Technology, Monash University and the Australian Astronomical Observatory. SkyMapper is owned and operated by The Australian National University's Research School of Astronomy and Astrophysics. The survey data were processed and provided by the SkyMapper Team at ANU. The SkyMapper node of the All-Sky Virtual Observatory (ASVO) is hosted at the National Computational Infrastructure (NCI). Development and support the SkyMapper node of the ASVO has been funded in part by Astronomy Australia Limited (AAL) and the Australian Government through the Commonwealth's Education Investment Fund (EIF) and National Collaborative Research Infrastructure Strategy (NCRIS), particularly the National eResearch Collaboration Tools and Resources (NeCTAR) and the Australian National Data Service Projects (ANDS). Parts of this project were conducted by the Australian Research Council Centre of Excellence for All-sky Astrophysics (CAASTRO), through project number CE110001020. I thank Christian Wolf and Chris Onken and also Alcione Mora of the ESA Gaia archive help desk for their help formulating and implementing database queries. I thank Duncan Forbes and my S5 colleagues for conversations about the halo and an anonymous referee for pointing out a misunderstanding. This work has made use of data from the ESA space mission Gaia, processed by the Gaia Data Processing and Analysis Consortium (DPAC). Funding for the DPAC has been provided by national institutions participating  in the Gaia Multilateral Agreement.

The Pan-STARRS1 Surveys (PS1) have been made possible through contributions of the Institute for Astronomy, the University of Hawaii, the Pan-STARRS Project Office, the Max-Planck Society and its participating institutes, the Max Planck Institute for Astronomy, Heidelberg and the Max Planck Institute for Extraterrestrial Physics, Garching, The Johns Hopkins University, Durham University, the University of Edinburgh, Queens University Belfast, the Harvard-Smithsonian Center for Astrophysics, the Las Cumbres Observatory Global Telescope Network Incorporated, the National Central University of Taiwan, the Space Telescope Science Institute, the National Aeronautics and Space Administration under Grant No. NNX08AR22G issued through the Planetary Science Division of the NASA Science Mission Directorate, the National Science Foundation under Grant No. AST-1238877, the University of Maryland, and Eotvos Lorand University (ELTE), the Los Alamos National Laboratory, and the Gordon and Betty Moore Foundation.
This publication makes use of data products from the Two Micron All Sky Survey, which is a joint project of the University of Massachusetts and the Infrared Processing and Analysis Center/California Institute of Technology, funded by the National Aeronautics and Space Administration and the National Science Foundation.

\section*{References}
Abraham, R. et al 2017, IAU Symposium, 321, 137\\
Arenou, F. et al 2018, A\&A, 616, 17\\
Brown, T. et al 2008, ApJ, 685, L121\\
Carollo, D. et al 2018, ApJ, 859, L7\\
Chen, S. et al 2018, ApJ, 867, 132\\
Chen, Y. et al 2018, ASP Conf. 514, 57\\
Ferguson, A. 2018, IAU Symposium 334, 43\\
Frogel, J., Persson, S.E. \& Cohen, J. 1980, ApJ, 239, 495\\
Girardi, L. et al 2004, A\&A, 422, 205\\
Girardi, L. 2016, AN, 337, 871\\
Helmi, A. et al 2018, Nature, 563, 43\\
Kaiser, N. et al 2010, SPIE 77330E\\
Lawrence, G. 2018, Honours thesis, Swinburne Univ.\\
Li, T. et al 2018, AAS 233 12904\\
Majewski, S. et al 1999, AJ, 118, 1709\\
Malhan, K. et al 2018, MNRAS, 481, 3442\\
Mould, J. 2014, PASA, 30, 27\\
Mould, J. et al 2018, PASA, 35, 1\\
Reid, I.N. et al 1990, ApJ, 348, 98\\
Renzini, A. 1977, Proc. Saas Fee Conf. {\it Advanced Stages in Stellar Evolution}, (Geneva: Geneva Obs)\\
Renzini,A. \& Buzzoni, A. 1986, in {\it The Spectral Evolution of Galaxies}, eds. C. Chiosi \& A. Renzini, Reidel Publishing, pp. 199). \\
Rizzi, L. et al 2007, ApJ, 661, 815\\
Rood, R. 1972, ApJ, 177, 681\\
Saremi, E. et al 2018, IAU Symposium 344, arxiv 1812.09725\\
Schlegel, D. et al 1998, ApJ, 500, 525\\
Schneider, D., Gunn, J. \& Hoessel, J. 1983, 264, 337\\
Skrutskie, M. et al 2006, AJ, 131, 1163\\
Shipp, N. et al 2018, ApJ, 862, 114\\
Torrealba, G. et al 2018, arxiv 1808.04082\\
Williams, B. et al 2017, ApJ, 846, 145\\
Wolf, C. et al 2018, PASA, 35, 010\\

\vfill\break
\setcounter{figure}{0}
\leftline{\bf Appendix}
Antlia 2 is a diffuse dwarf galaxy discovered using a Gaia kinematic filter (Torrealba et al 2018).
Such a filter can be added to a query such as the one in  $\S$2.1 and for a radius of 0.63 degrees around the centre of Antlia 2, requiring $\pi~<$ 0.1 mas, PMRA(cos$\delta$) $>$ --1.5 mas/yr and $|$PMDEC$| ~<$ 1.5 mas/yr
yields the CMD in Figure A1.
\begin{figure}
\begin{center}
\includegraphics[width=1.2\columnwidth,angle=-90]{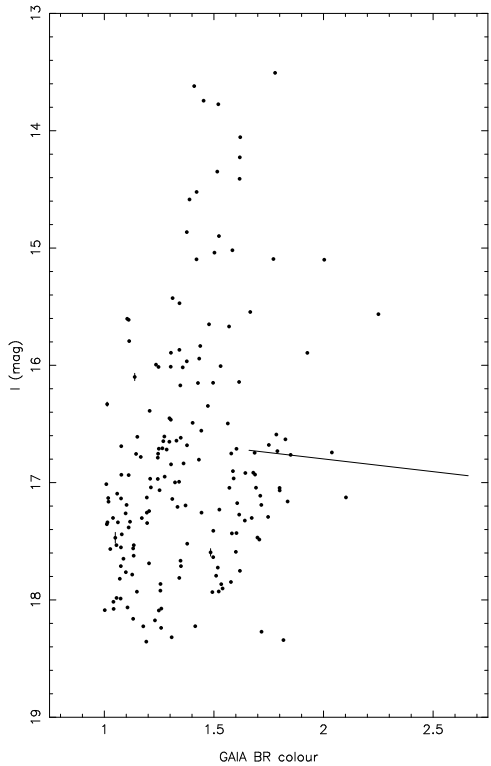}
\caption{Kinematically filtered CMD for Antlia 2 with close to zero parallax and proper motion.
The I mag is on the Cousins system.
The sloping line is the TRGB of Rizzi et al (2007), moved to apparent distance modulus 20.5 plus A$_I$ = 0.3 mag. A few stars rise above that TRGB in luminosity.}
\end{center}
\end{figure} 
The reddening of Antlia 2 is E(B-V) = 0.16 mag according to Schlegel et al (1998). The red giant branch is clearly seen in the SkyMapper-Gaia photometry and is not seen in a control sample annulus 1 degree from Antlia 2. Evidently,
Antlia 2 is closer to 110 kpc away than the 129.4 $\pm$ 6.5 kpc measured by Torrealba et al.  
\end{document}